\documentclass[12pt]{article}

\usepackage[latin1]{inputenc}
\usepackage{graphicx, color}
\usepackage{hyperref}
\definecolor{link}{rgb}{0.1,0.1,0.9}
\hypersetup{colorlinks=true,linkcolor=link,citecolor=link,urlcolor=link,linktocpage}
\usepackage{epstopdf}
\usepackage{color}
\usepackage{amsmath}
\usepackage{amssymb}
\usepackage{longtable}
\usepackage{mathtools}

\usepackage{scicite}

\usepackage{times}

\graphicspath{{F:/HarishWork/SciAdvance/}}

\newenvironment{sciabstract}{%
\begin{quote} \bf}
{\end{quote}}

\title{Multiple phases with tricritical point and Lifshitz point in skyrmion host Cu$_2$OSeO$_3$}

\author
{Harish Chandr Chauhan$^{\ast}$, Birendra Kumar, Jeetendra Kumar Tiwari, Subhasis Ghosh\\
\\
\normalsize{$^{\ast}$School of Physical Sciences, Jawaharlal Nehru University, New Delhi-110067, India}\\
}

\date{}

\begin{document} 


\baselineskip24pt


\maketitle


\begin{sciabstract}
  Magnetic skyrmions, a topologically stable spin swirling object, have attracted a great interest due to their potential applications in future spintronics and ultra high dense magnetic memory devices. Cu$_2$OSeO$_3$ is only known insulating chiral helimagnet with multiple phases including skyrmion phase. Existence of skyrmion phase as well as first and second order phase transitions make Cu$_2$OSeO$_3$ a promissing candidate for investigating complex magnetic phenomena. Here, we report that both first and second order magnetic phase transitions are responsible in determining the phase diagram with atleast two multicritical point in Cu$_2$OSeO$_3$. Fluctuation-induced first order transition is realized as a precursor for skyrmion phase over a small window of temperature of magnetic field. The evolution of field dependent entropy at the phase transition provides the evidence for tricritical point. Furthermore, existence of commensurate and incommensurate phases, alongwith co-existence of three second order phase transitions provide evidence for the existence of Lifshitz point.
\end{sciabstract}

\section*{Introduction}
Skyrmion, a topologically nontrivial spin-swirling object that can be treated as a particle of nanometre size has been observed in MnSi \cite{1,2}, FeGe \cite{3,4,5}, Fe$_{1-x}$Co$_x$Si \cite{6,7}, Cu$_2$OSeO$_3$ \cite{8,9,10} with cubic B20 structure. Skyrmion has also been observed in metallic thin films and multilayer \cite{11}. These skyrmionic materials have attracted great interest due to their novel properties and potential applications in future spintronics and magnetic memory devices. Cu$_2$OSeO$_3$ crystallizes in the space group P$2_13$, same as in MnSi, FeGe,  Fe$_{1-x}$Co$_x$Si and other B20 structures. Cu$_2$OSeO$_3$ has broken inversion symmetry and three-fold rotational symmetry along [111]. Cu$_2$OSeO$_3$ is possibly the only known insulating \cite{8} chiral magnetic material \cite{12,13,14,15}. Seki \textit{et al} \cite{8} have first observed the skyrmion phase in Cu$_2$OSeO$_3$ and shown that there are two inequivalent sites of Cu$^{2+}$ ions surrounded by two types of CuO$_5$ polyhedra, one with square pyramidal and other with trigonal bipyramidal structure, in the ratio of 3:1. These two sublattices of Cu$^{2+}$ ions are  responsible for local ferrimagnetic ordering as well as helimagnetic phase below T$_C$ in this material \cite{12, 14,16}. Due to the broken inversion symmetry and strong spin$-$orbit coupling (SOC) in Cu$_2$OSeO$_3$, there are two contributions in spin exchange interaction, (i)  $\vec{S}_i$$\cdot$$\vec{S}_j$, symmetric Heisenberg exchange interaction (HEI) and (ii) $\vec{D}_{ij}$$\cdot$$\vec{S}_i$$\times$$\vec{S}_j$, anti-symmetric Dzyloshinskii-Moria exchange interaction (DMEI) \cite{17, 18} where $\vec{S}_i$, $\vec{S}_j$ are the spins at  $i^{th}$ and $j^{th}$ sites respectively and $\vec{D}_{ij}$ is the Dzyloshinskii-Moria vector joining $i^{th}$ and $j^{th}$ site. DMEI is the result of HEI with strong SOC and broken inversion symmetry. The competition between DMEI and HEI stabilizes the ground state of Cu$_2$OSeO$_3$ in helical spin texture with fixed handedness (spin chirality) below a transition temperature ($T_c$) of 58 K \cite{8,16}. A weak external field overcomes this helical phase and orient it along the direction of field leading to conical phase \cite{19} as well as skyrmion phase also known as A-phase just below $T_c$. In higher field, HEI dominates over DMEI and induces conical phase into field polarized (FP) phase \cite{8,16}. Small angle neutron scattering (SANS) \cite{9} and Lorentz transmission electron microscopy (LTEM) \cite{8,10} results reflect multiple phases in Cu$_2$OSeO$_3$ just below T$_C$, however these results are still inconclusive to resolve several issues such as: (i) is the phase diagram of all skyrmionic materials generic?, (ii) can SANS and LTEM identify all the phases?, (iii) as both first and second-order phase transition exist in these materials, then where is the tricritical point (TCP)? (iv) is there a precursor phase required for the transition from paramagnetic (PM) to helimagnetic(HM) phase?, (v) how have the phase boundaries been constructed? Resolving these issues is complicated by the fact that the skyrmion phase is thermodynamically stable only in a small pocket of phase space.

To realize the above concerned problems, we did high precession magnetic measurements on Cu$_2$OSeO$_3$ over wide range of magnetic field and temperature. From our observations we have been able to present (i) the different phases present in Cu$_2$OSeO$_3$ near $T_c$, (ii) the order of phase transitions between different phases, (iii) existence of a precursor phase required for the transition from PM to HM phase, (iv) existence of TCP and a special multicritical point known as Lifshitz point (LP), and (v) the exact transition lines between phases. 

\section*{Results and discussions}

Reitvield refinement of the XRD data of Cu$_2$OSeO$_3$ (Fig. 1a) obtained by  Rigaku Miniflex 600 X-Ray Diffractometer with Cu$-$K$_\alpha$ radiation.  It confirms the formation of single cubic phase of Cu$_2$OSeO$_3$ in the space group P$2_13$ with lattice constant  a = b = c = 8.922 $\AA$  and $ \alpha = \beta = \gamma = 90^0$. \cite{20}. Figure 1b is the unit cell of Cu$_2$OSeO$_3$ with broken inversion symmetry and three fold rotational symmetry along [111]. It gives two CuO$_5$ polyhedra: one is square pyramidal (Fig. 2c) and other in trigonal bi-pyramidal (Fig. 2d) in 3:1 ratio respectively. This two unequal ratio polyhedra's are responsible for two sub-lattices giving rise to local ferrimagnetic ordering in Cu$_2$OSeO$_3$.
\begin{figure}
	\includegraphics[width=15.8cm]{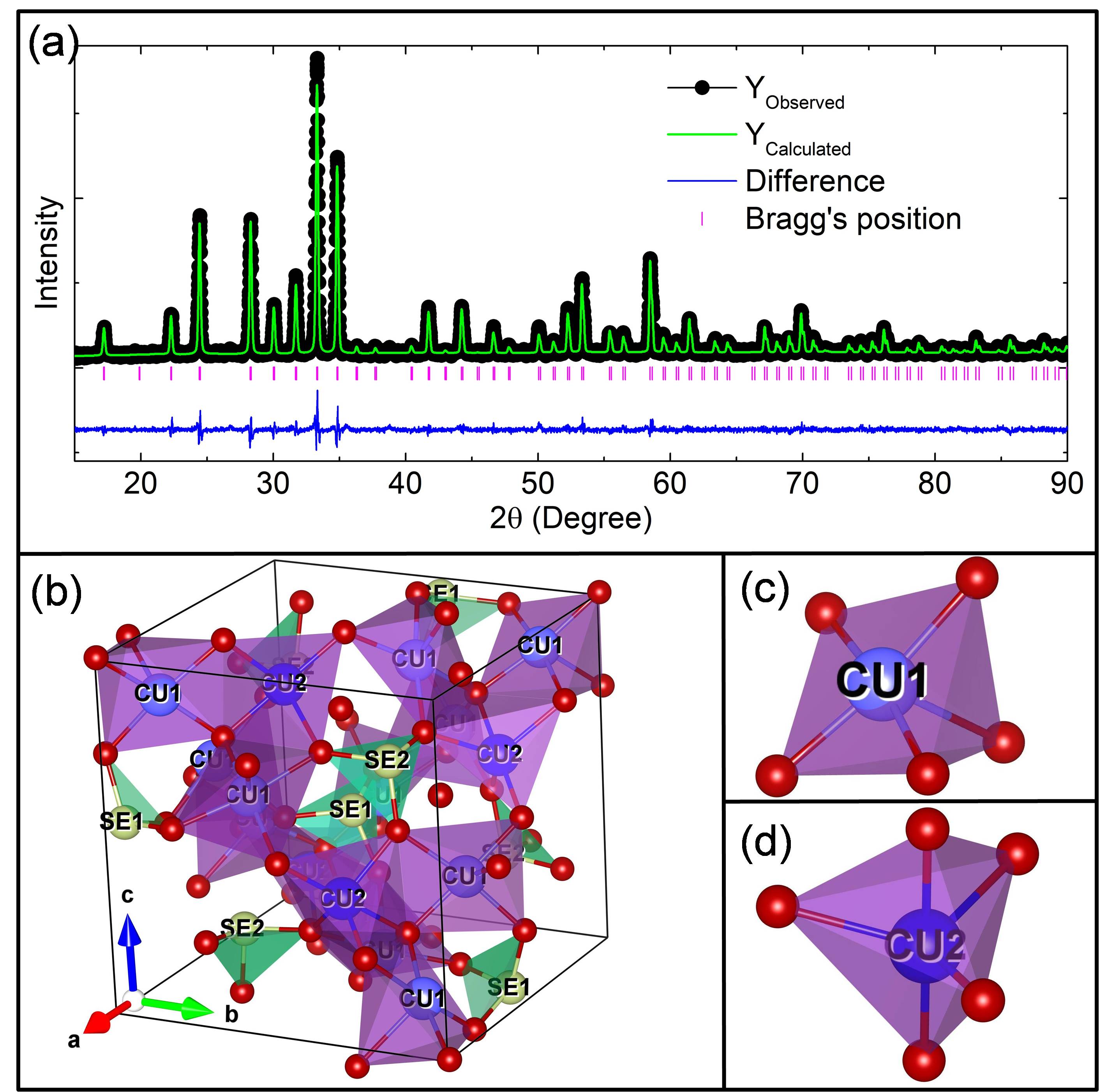}
	\caption{(a) Reitveld Refined XRD pattern of Cu$_2$OSeO$_3$, (b) Unit cell of Cu$_2$OSeO$_3$, (c) Square pyramidal CuO$_5$, and (d) Triginal bi-pyramidal CuO$_5$ polyhedra.}
\end{figure}

\begin{figure}
	\includegraphics[width=15.8cm]{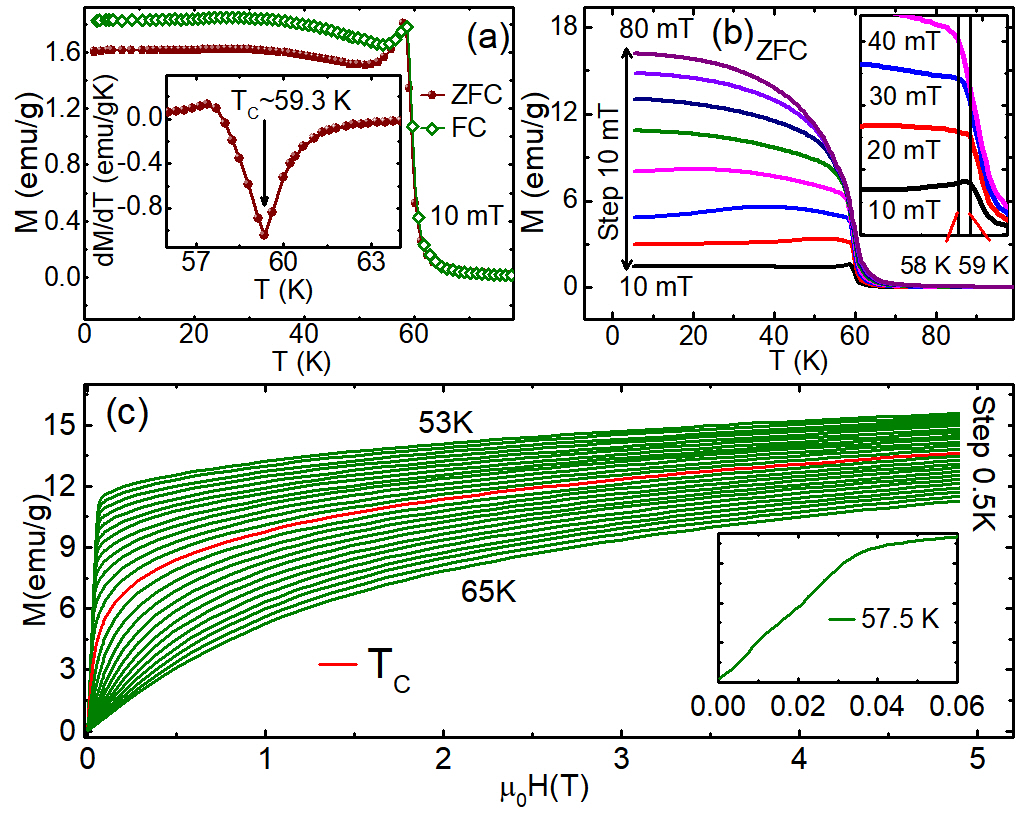}
	\caption{(a) Field cooled (FC) and zero field cooled (ZFC) data of Cu$_2$OSeO$_3$ taken at 10 mT. Inset of (a) is the first derivative of the ZFC data. The minimum gives $T_c$ $=$ 59.3 K. (b) ZFC of Cu$_2$OSeO$_3$ from 10 mT to 80 mT with increment step of 10 mT. The variation in the ZFC curves is the clear indication of helimagnetic to field polarised phase transition around 40 mT. The inset shows the magnified view till 40 mT from 50 K to 62 K. (c) Field dependent magnetization (M$-$H) data of Cu$_2$OSeO$_3$ at various temperatures from 53 K to 65 K with increment step of 0.5 K. No saturation in the M$-$H data implies non ferromagnetic Cu$_2$OSeO$_3$.}
\end{figure}
Field-cooled (FC) and zero field-cooled (ZFC) curves (Fig. 2a) taken at 10 mT shows the bifurcation around 59 K which indicates that the material is paramagnetic above 59 K and HM below 59 K. First derivative of the ZFC curve gives transition temperature, T$_C$ = 59.3 K at 10 mT.  ZFC curves taken at different applied fields (Fig. 2b) shows transition of HM phase into field polarised (FP) phase above 40 mT. The variation in the ZFC curves are the indication of multiphase existence in Cu$_2$OSeO$_3$. For further study of the concerned problems, we took field dependent magnetization (M$-$H) data upto 5 T at various temperatures ranging from 53 K to 65 K with incement step of 0.5 K as shown in Fig. 2c. No saturation in the magnetization curves is the clear indication that Cu$_2$OSeO$_3$ is non feromagnetic. The inset of Fig. 2c is the M$-$H data at 57.5 K which shows step like variation in magnetic mament. This is again the indication of multiphase existence in Cu$_2$OSeO$_3$ below T$_C$.

The magnetostatic free energy based on Ginzburg's criteria as a function of the order parameter (magnetization, $M$) upto $M^4$  can be reduced to Arrot relation \cite{21} as: 
\begin{equation}
M^2 = a\left( \frac{\mu_0H}{M} \right)+b.
\end{equation}
\begin{figure}
	\includegraphics[width=15.8cm]{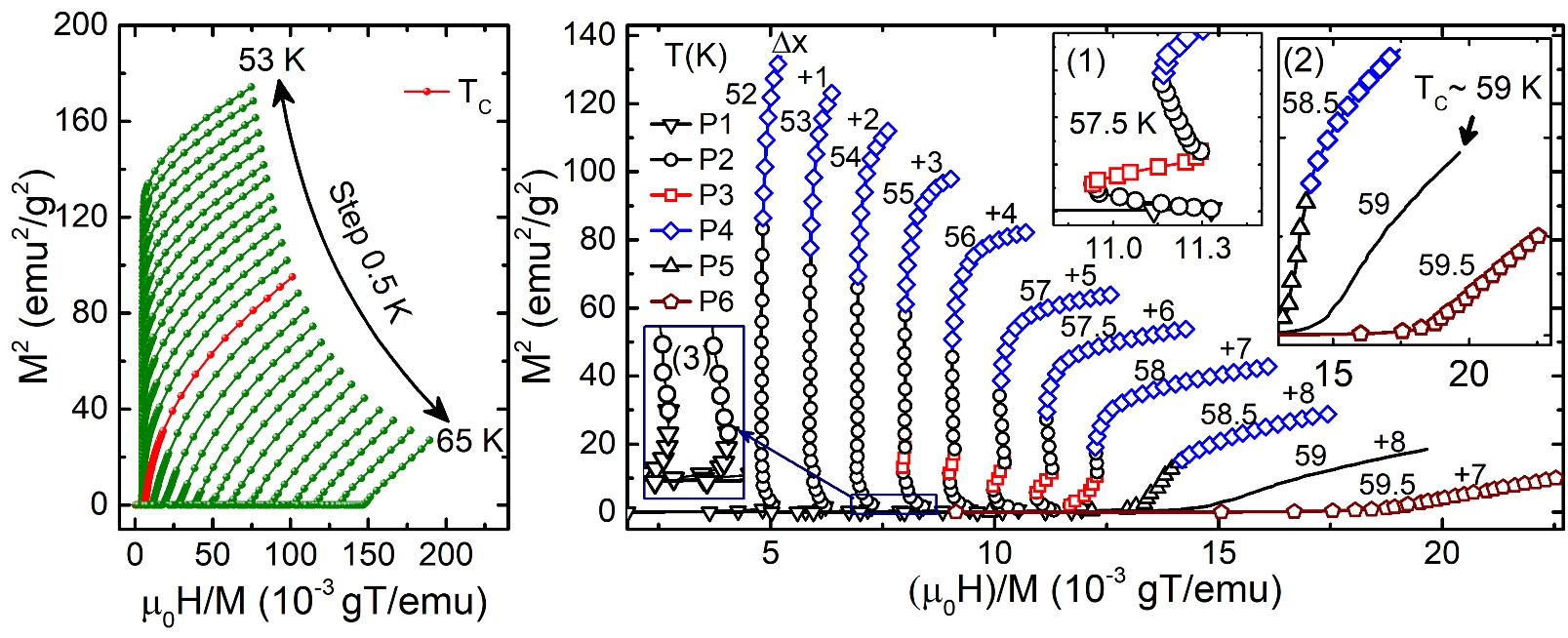}
	\caption{(a) The Arrot plot of the magnetization data upto 1 T. The nonlinear variation of the Arrot plot denies the ferromagnetic nature of Cu$_2$OSeO$_3$. (b)The Arrot plot of magnetization data up to 60 mT from 52 K to 58 K and up to 50 mT above 58 K. The different temperature curves have been moved horizontally to see the clear variation.  The different colors represent the phases present in Cu$_2$OSeO$_3$. P1 (down triangles): MDH phase, P2 (circles): SDC phase, P3 (squares): skyrmions (A-phase), P4 (diamonds): field polarized phase, P5 (up triangles) represents the fluctuation disordered (FD) phase surrounded by SDC phase, A-phase and PM phase, and P6 (pentagons) represents the PM phase. Inset (1) is the Arrot plot at 57.5 K in the field range of 3 mT to 35 mT and inset (2) is the magnified view of the Arrot plot at 58.5 K, 59 K and 59.5 K. Inset (3) is the extended view in the vicinity of MDH and SDC phase.}
\end{figure}
where $a$ and $b$ are temperature and magnetic field dependent constants. Arrot plots of the M$-$H data (Fig. 3a) are not collinear which implies existence of other phase (HM phase) than ferromagnetic. The reported phase diagram is at very low field and below T$_C$. So, we took Arrot plot upto 60 mT (Fig. 3b) at the temperatures ranging from 52K to 59.5K.  The Arrott data are moved horizontaly to see the clear variation at various temperatures.  Inset (1) of Fig.3b shows Arrot plot at 57.5 K from 3 mT to 35 mT. It can be seen that there are multiple positive and negative slopes of Arrot plot below $T_c$. The change in the slope of the Arrot plot is the indication of field induced phase transition in  Cu$_2$OSeO$_3$. Inset (2) of Fig.3b shows the change in the curvature of Arrot plot from concave to convex \cite{22} around 59 K which is same as the $T_c$ obtained from ZFC (Fig.2b) indicating that the PM phase is above 59 K. As different slopes represent different phases, phases shown by P1(down triangles), P2(circles),  P3 (squares) and P4 (diamonds) are for multidomain helical (MDH) \cite{8}, single domain conical (SDC) \cite{1}, skyrmion (A-phase) \cite{1, 2, 3, 4, 5, 6, 7, 8, 9, 10} and field polarised (FP) phase, respectively. Our result is in agreement with the results obtained from SANS \cite{9,16} and LTEM \cite{8, 10}. All three phases: MDH, SDC and A-phase have the boundary in the vicinity of  $T_c^\prime=58$ K (Fig.3b) which is not the Curie temperature for transition to PM phase as that exists above $59.3$ K (Fig.2b) Hence there must be another phase shown by P5 (up triangles), between $T_c^\prime$ and $T_c$ . The finite moment (see Fig.2c) in this region confirms the existence of DMI which is essential as a precursor for the transition from PM to HM state. It has been shown theoretically \cite{23} that fluctuation induced inhomogeneous phase with finite moment is required as precursor for transition to HM state in case of first order magnetic transition. Similar fluctuation disordered (FD) phase may have been observed in MnSi \cite{24} which  is characterized by strongly interacting chiral fluctuation that induces first order Brazovskii transition \cite{25, 26, 27}. Also, according to Banerjee's criteria \cite{28}, $a>0$ in Eq. (1) correspond to second order phase transition and $a<0$ correspond to first order phase transition at $T_c$. Thus the transitions from FP to PM and FD to PM phase are second order.

\begin{figure}
	\includegraphics[width=15.8 cm]{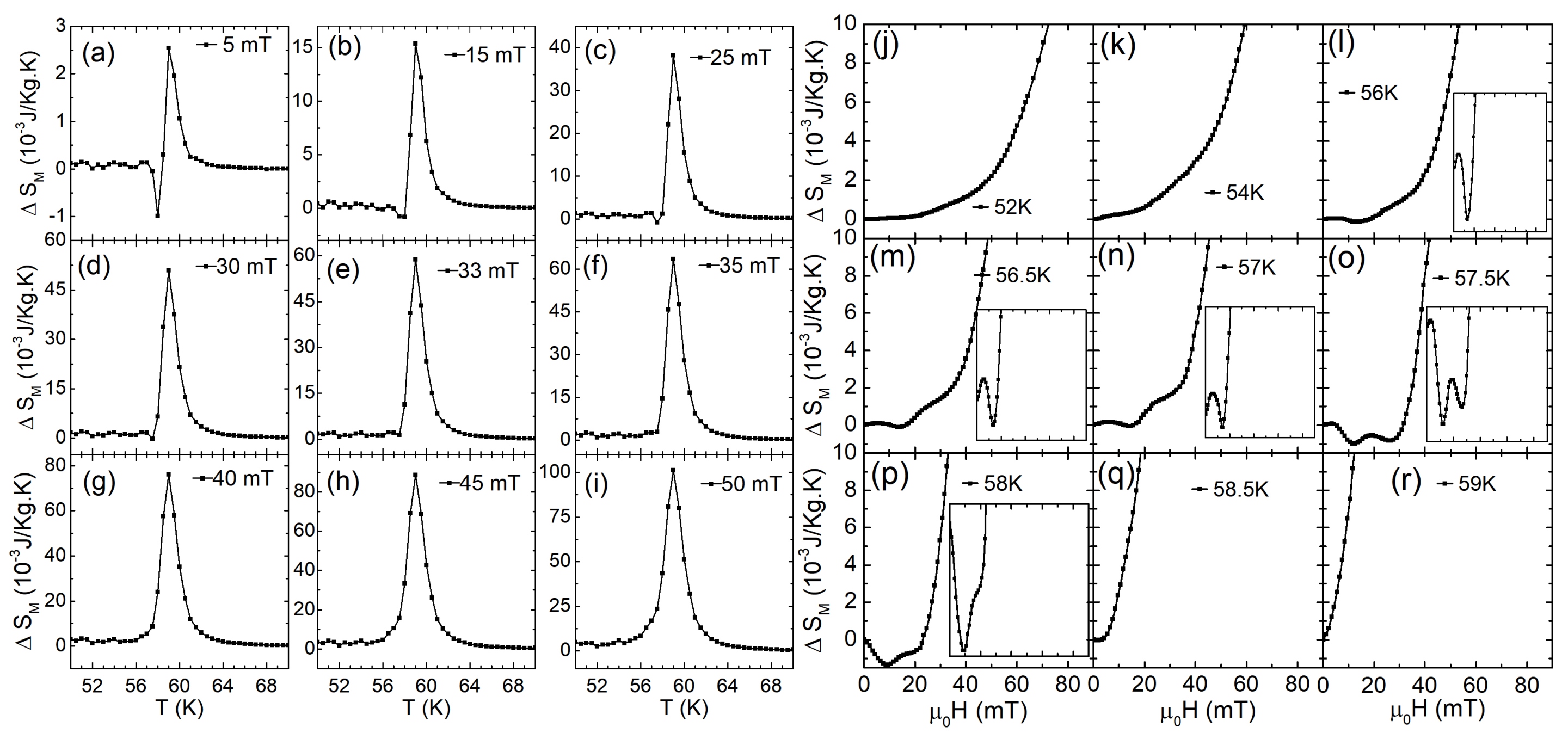}
	\caption{(a)$-$(i) Change in magnetic entropy, $\Delta S_M$ as a function of temperature at different applied magnetic fields varying from 5 mT to 50 mT. The position of the peak observed near 58.07 K at 5 mT as shown in (a) changes with applied field and disappears at 57.5 K and 33 mT as shown in (e). A discontinuous change in slopes of transition curves of $\Delta S_M$ is observed around 59 K (a$-$b) below 15 mT. This could be due to insufficient data points. $\Delta S_M$ (c$-$i) follows a continuous change in transition curves around 59 K and above 15 mT. j$-$r is the variation of $\Delta S_M$ as a function of applied field at different temperatures. The insets of the graphs are the magnified view near the transition field.}
\end{figure}

Using Gibbs free energy expression and Maxwell's relation one can compute the change in entropy, $\Delta S_M$, as:

\begin{equation}
\Delta S_M = \int_{\mu_0H_1}^{\mu_0H_2}\left( \frac{\partial M}{\partial T} \right)_{\mu_0H} d(\mu_0H).
\end{equation}

The magnetic entropy change, $\Delta S_M$, in Cu$_2$OSeO$_3$ around $T_c$ can be  investigated by using the M$-$H curves at various temperatures \cite{29}.
Figure 4  shows the variation of $\Delta S_M$ of Cu$_2$OSeO$_3$ with temperature at different applied magnetic fields (Figs. 4a$-$i) and with field at different temperatures (Figs. 4j$-$r). The maximum change in $\Delta S_M$ is observed at $T_c$. The magnitude of change in magnetic entropy increases gradually with applied field. $\Delta S_M$ shows a small but clear depth at  $T_C^\prime$ which changes its position continuously from 58.07 K at 5 mT (Fig. 4a) to 57.5 K at 33 mT (Fig. 4e). This feature, which is absent in all previous work \cite{8,9,10,15,16,19}, has been observed between $T_c^\prime$ and $T_c$, may be of similar origin  behind FD regime reported in \cite{24}. The magnitude of the depth at $T_c^\prime$ first decreases up to 11 mT then increases from 11 mT to 16 mT and again decreases from 16 mT to 23 mT. Above  23 mT, the depth at $T_c^\prime$ starts increasing and vanishes near 33 mT. It is observed (Figs. 4a$-$e) that $\Delta S_M$ starts decreasing when approaches $T_c^\prime$ and then increases discontinuously above $T_c^\prime$. This sudden change in $\Delta S_M$ implies that the system undergoes through a first order phase transition at  $T_c^\prime$ \cite{30}. The variation in $\Delta S_M$ is quite fast around $T_c$ till 15 mT (Fig. 4a,b) and continuous above 15 mT (Figs. 4c$-$i) leading second order phase transition \cite{30}. This will violate the Banerjee's criteria if the transition is first order below 15 mT. Thus the sharp change could be due to insufficient data points near $T_c$  or due to the effect of first order transition at $T_c^\prime$ because $T_c$ is nearly 1 K away from $T_c^\prime$.

A small but significant discontinuous change in $\Delta S_M$ has been observed at the transition from SDC to A-phase as shown in the inset of Figs. 4l$-$p. The change in entropy from A-phase to SDC phase at the upper boundary is not observed clearly. Since there are only two phases (SDC and FD regime) encompassing the A-phase, the transition from A-phase to SDC phase will be a weak first order transition \cite{24}. It was observed (Figs. 4j$-$o) that variation of $\Delta S_M$ is continuous from MDH to SDC and from SDC to FP phases which are the confirmation of second order phase transitions. Figure 4o shows the discontinuous transition from A-phase to SDC phase at 57.5 K. This is again the evidence for the first order phase transition from A-phase to SDC phase (dip near 12 mT) and from SDC to FD regime (dip near 28 mT). The feature of change in entropy is observable till 58 K (Fig. 4q). At 59 K, change in entropy is maximum after that it starts decreasing. This is again the evidence for the existence of FD regime between $T_c^\prime$ and $T_c$.
\begin{figure}
	\includegraphics[width=15 cm]{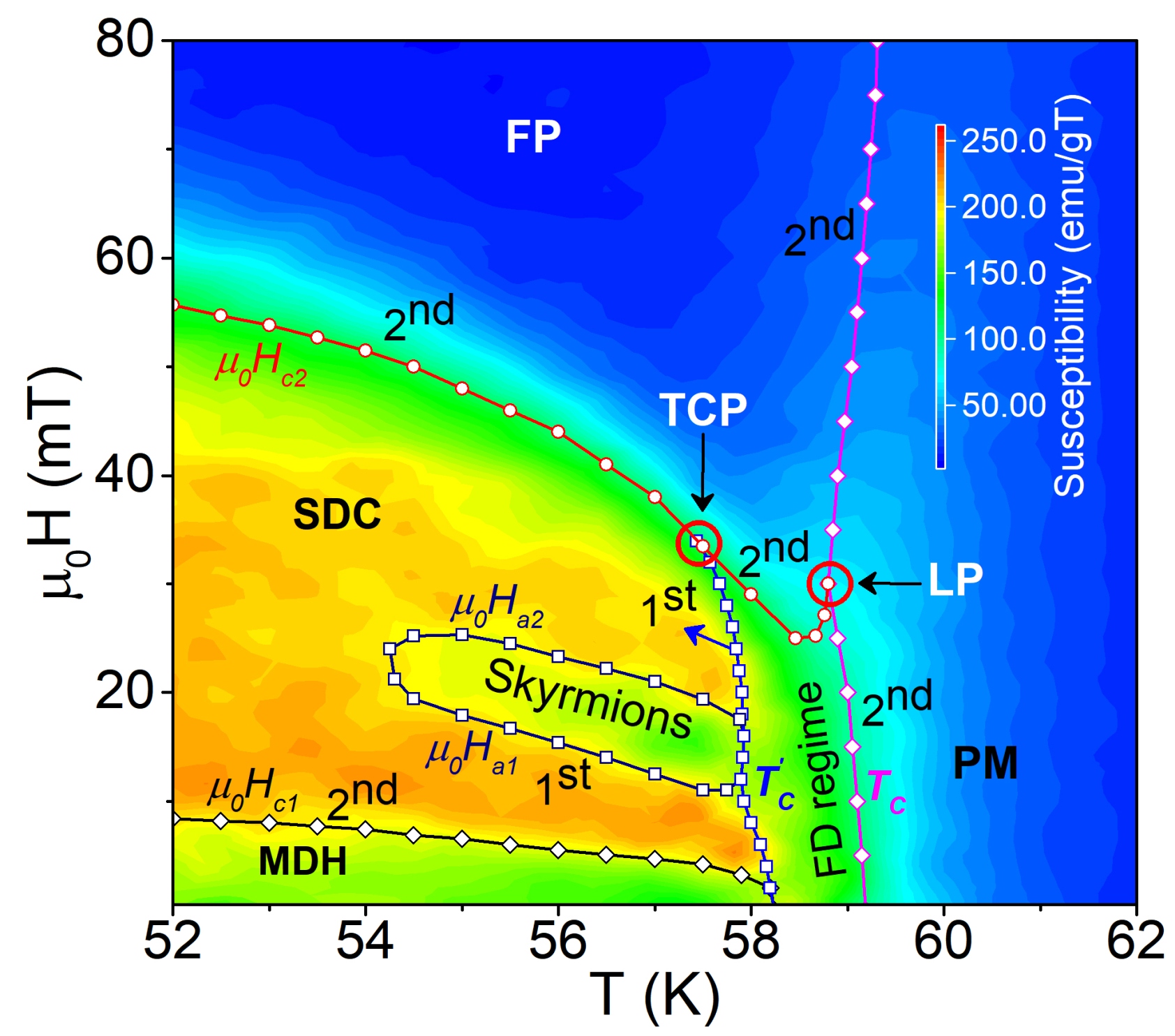}
	\caption{\textbf{Schematic phase diagram of Cu$_2$OSeO$_3$:} The magnetic phase diagram of Cu$_2$OSeO$_3$ derived from the Arrot plot and entropy analysis. Below $T_c$, MDH phase, SDC phase, Skyrmion (A-phase) and field polarized phase are observed. FD phase is just below $T_c$ in the width of around 1 K which vanishes at 57.5 K near 33 mT. PM phase is above $T_c$.}
\end{figure}

The phase diagram of Cu$_2$OSeO$_3$ with all possible phases have been constructed from the temperature and field dependent susceptibility data obtained from M$-$H data is shown in Fig. 5. We defined the phase boundaries using various techniques:(i) Arrot analysis (Fig. 3) to draw the MDH to SDC ($\mu_0H_{c1}$) and SDC to A-phase boundaries ($\mu_0H_{a1}$ and $\mu_0H_{a2}$), (ii) the change in entropy analysis (Fig. 4) to draw $T_c^\prime$ connecting MDH, SDC and A-phase to FD regime, (iii)  first derivative of the temperature dependent magnetization curves obtained from the M$-$H curves taken at various temperatures to draw $T_c$ connecting FD regime and FP phase to PM phase, and (iv) scaling of the susceptibility curves obtained from the M$-$H data to construct the boundary of FP phase ($\mu_0H_{c2}$) connecting SDC and FD regime. $\mu_0H_{c2}$ is the transition line joining the point of inflection of susceptibility curve at different temperatures.

In the phase diagram (Fig. 5), the MDH phase lies below $\mu_0H_{c1}$. The transition lines $\mu_0H_{c1}$, $\mu_0H_{c2}$, $\mu_0H_{a1}$, $\mu_0H_{a2}$ and $T_c^\prime$ cover the SDC phase. Skyrmion phase is in the temperature range from 54 K to 58 K and in the field range from 12 mT to 28 mT. FP phase is above $\mu_0H_{c2}$ and below $T_c$. The PM phase is above $T_c$. $T_c^\prime$, $\mu_0H_{c2}$ and $T_c$ surround FD regime. As seen clearly that the first order transition line, $T_c^\prime$, becomes second order ($\mu_0H_{c2}$) at (57.5 K, 33 mT) where  SDC, FD regime and FP phases meet. These observations suggest that (57.5 K, 33 mT) must be a TCP. Further, the collinear spins are responsible for the FP phase as a commensurate phase while non-collinear spins make HM phase as an incommensurate phase \cite{31}. The presence of  anti-symmetric DMEI \cite{32} is responsible for FD regime as an incommensurate phase. Also the transition line, $\mu_0H_{c2}$ meets $T_c$ tangentially at (58.8 K, 30 mT) joining three phases viz FP, FD regime and PM phase. Based on these observations the point (58.8 K, 30 mT) in the phase diagram can be interpreted as Lifshitz point (LP) which was first proposed by Honreich, Luban and Shtrikman \cite{33} as a special multicritical point with an incommensurate phase joining all three phases with a common tangent. Similar LP have been observed in UPd$_2$Si$_2$ \cite{31}, UAs$_{1-x}$Se$_x$\cite{34}, LaCrGe$_3$ \cite{35} and a variety of different systems \cite{31, 32, 36, 37}.

In conclusion, we have constructed the complete phase diagram of Cu$_2$OSeO$_3$ near skyrmion phase with the help of Arrot plot \cite{22, 29} and entropy analysis \cite{30, 31}. The exact transition line from PM to other phases was missing link in the previously reported phase diagrams \cite{8,9, 10, 15, 16, 19}. We have investigated the order of phase transitions among the six phases present in Cu$_2$OSeO$_3$ near transition temperature, $T_c$. SANS and LTEM cannot detect the FD regime which can only be observed by high precession magnetic measurement. It is found that FD regime is an essential phase for the transition from PM to HM phase in those skyrmionic materials which have non-centrosymmetry and cubic B20 structure. A TCP has been observed at (57.5 K, 33 mT) where first order transition line, $T_c^\prime$, meets the second order transition line, $\mu_0H_{c2}$. The closer analysis of the phase diagram confirms the existence of Lifshitz point at (58.8 K, 30 mT).



\section*{Acknowledgments}
 We thank AIRF-JNU for providing facilities for PPMS and XRD measurement. H.C.C. acknowledges UGC-CSIR for the financial support through fellowship. This project is  partially supported by DST-PURSE.

\section*{Methods}
	Polycrystalline Cu$_2$OSeO$_3$ were grown by standard solid-state reaction method. A homogeneous mixture of high purity CuO and SeO$_2$ powder in the molar ratio 2:1 was taken. The pellets formed from this mixture were then sealed in an evacuated quartz tube. The quartz ampule was placed horizontally into a muffle furnace and heated to 600 $^0$C with 50 $^0$C$/$hr. We have annealed the sample for 5 weeks by holding the quartz ampule at 600 $^0$C. 
	
	\subsection*{X-ray diffraction (XRD).} To confirm the cubic phase of  Cu$_2$OSeO$_3$, we performed the powder x-ray diffraction (XRD). The room temperature XRD data were collected using Rigaku Miniflex 600 X-Ray Diffractometer with Cu$-$K$_\alpha$ radiation. Further, rietveld refinement using Full Proof suite software was performed on the XRD data. It confirmed the formation of the single cubic phase of Cu$_2$OSeO$_3$.
	
	\subsection*{Magnetic measurements.} The magnetic measurements were done using physical properties measurement system (PPMS) by three different methods as follows: 
	
	\textbf{(i)  Field cooled (FC) temperature scans}: In the presence of magnetic field the Cu$_2$OSeO$_3$ sample was cooled from room temperature to the desired low temperature. Finally the temperature dependent magnetization data was recorded during heating. 
	
	\textbf{(ii) Zero-FC (ZFC) temperature scans}: The sample was brought at low temperature and then by applying magnetic field the temperature dependent magnetization data was recorded during heating.
	
	\textbf{(iii) ZFC magnetic field scans}: The sample was brought at the various required temperatures and held untill the thermal equilibrium was reached. The field dependent magnetization data was recorded. First quadrant magnetization data have been taken with the step of 1 mT from zero field to 200 mT for exact analysis of phase diagram and associated phase transitions between phases. Above 200 mT the step size was increased to 100 mT and data was taken up to 1 T to analyze the magnetic properties of Cu$_2$OSeO$_3$.
	
	\bibliographystyle{Science}
	\bibliography{RefSA}

\end{document}